\documentclass
[preprint,tightenlines,showpacs,showkeys,floatfix,nofootinbib,superscriptaddress,fleqn]{revtex4}
\usepackage{amssymb}
\usepackage{graphicx}
\usepackage{bm}
\usepackage{longtable}
\usepackage{amsmath}
\usepackage{amsfonts}
\begin{document}
\preprint{PNU-NTG-01/2007}
\preprint{PNU-NURI-01/2007}
\title{Quark spin content of the proton,  hyperon semileptonic decays,
and the decay width of the $\Theta^{+}$ pentaquark}
\author{Ghil-Seok Yang}
\email{yangg@tp2.rub.de}
\affiliation{Institut f\"ur Theoretische Physik II, Ruhr-Universit\" at Bochum, D--44780
Bochum, Germany}
\author{Hyun-Chul Kim}
\email{hchkim@pusan.ac.kr}
\affiliation{Department of Physics, and Nuclear Physics \& Radiation Technology Institute
(NuRI), Pusan National University, 609-735 Busan, Republic of Korea}
\author{Klaus Goeke}
\email{Klaus.Goeke@rub.de} \affiliation{Institut f\"ur
Theoretische Physik II, Ruhr-Universit\" at Bochum, D--44780
Bochum, Germany}
\date{March 2007}
\begin{abstract}
~~~\newline Using the exisiting experimental data for hyperon
semileptonic decays and the flavor-singlet axial-vector charge
$g_{A}^{(0)}$ from polarized deep inelastic scattering of the
proton, we derive the decay width of the $\Theta^{+}$ pentaquark
baryon. We take into account the effects of flavor SU(3) symmetry
breaking within the framework of the chiral quark-soliton model.
All dynamical parameters of the model are fixed by using the five
experimental hyperon semileptonic decay constants and flavor
singlet axial-vector charge. We obtain the numerical results of
the decay width of the $\Theta^{+}$ pentaquark baryon as a
function of the pion-nucleon sigma term $\Sigma_{\pi N}$ and
investigate the dependence of the decay width of the $\Theta^{+}$
on the $g_{A}^{(0)}$, varying the $g_{A}^{(0)}$ within the range
of the experimental uncertainty. We demonstrate that the combined
values of all known semileptonic decays with the generally
accepted value of $g_{A}^{(0)} \approx 0.3$ for the proton are
compatible with a small decay width $\Gamma_{\Theta KN}$ of the
$\Theta^{+}$ pentaquark, i.e. $\Gamma_{\Theta KN} \leq 1$~MeV.

\end{abstract}

\maketitle

\section{Introduction}
Since Diakonov, Petrov and Polyakov~\cite{Diakonov:1997mm}
predicted in the chiral quark-soliton model ($\chi$QSM) the mass
and narrow decay width of the pentaquark baryon $\Theta^{+}$ with
strange quantum number $S=+1$ and leading quark Fock structure
$uudd\bar{s}$, there has been an enormous amount of theoretical
and experimental works (see, for example, recent reviews for the
experimental results~\cite{Hicks:2005gp} and for the theoretical
investigations~\cite{Zhu:2004xa,Goeke:2004ht,Oh:2004wp}). Note
that there is an earlier prediction by Prasza{\l }owicz of the
mass in the soliton approach of the Skyrme
model~\cite{Praszalowicz}. Many experiments have announced the
existence of the $\Theta^{+}$ after the first independent
observations by the LEPS~\cite{Nakano:2003qx} and
DIANA~\cite{Barmin:2003vv} collaborations,
 while the $\Theta^{+}$ has not been seen in almost all high-energy experiments.
Moreover, an exotic $\Xi_{\overline{10}}$ state was observed by
the NA49 experiment at CERN~\cite{Alt:2003vb}, though its
existence is still under debate.

A very recent CLAS experiment dedicated to search for the
$\Theta^{+}$ has announced null results of finding the
$\Theta^{+}$ in the reaction $\gamma p\rightarrow
\bar{K}^{0}\Theta^{+}$~\cite{Battaglieri:2005er}. The subsequent
experiment has also not found any evidence for the $\Theta^{+}$ in
$\gamma d\rightarrow pK^{-}\Theta^{+}$~\cite{McKinnon:2006zv}.
Though these experiments are the measurements with high
statistics, it is too early to conclude the absence of the
$\Theta^{+}$. Note e.g. that the DIANA collaboration has continued
to search for the $\Theta^{+}$ and found the formation of a narrow
$pK^{0\text{ }}$peak with mass of $1537\pm2$ MeV/$c^{2}$ and width
of $\Gamma=0.36\pm0.11$ MeV in the $K^{+}n\rightarrow K^{0}p$
reaction~\cite{Barmin:2006we}.  Moreover, several new experiments
searching for the $\Theta^{+}$ are in
progress~\cite{Hotta:2005rh,Miwa:2006if}. In this obscure status
for the $\Theta^{+}$, more efforts are required for understanding
the $\Theta^{+}$ theoretically as well as experimentally. In
addition, a recent GRAAL experiment~\cite{Kuznetsov:2004gy}
announced the evidence of a new nucleon-like resonance with a
seemingly narrow decay width $\sim10$ MeV and a mass $\sim1675$
MeV in the $\eta$-photoproduction from the neutron target. This
new nucleon-like resonance, $N^{\ast}(1675)$, may be regarded as a
non-strange pentaquark because of its narrow decay width and
dominant excitation on the neutron target which are known to be
characteristic for typical pentaquark
baryons~\cite{Polyakov:2003dx}, though one should not exclude a
possibility that it might be one of the already known $\pi N$
resonances (possibly, $D_{15}$)~\cite{Kuznetsov:2006de}. This
GRAAL data is consistent with the results for the transition
magnetic moments in the chiral quark-soliton model
($\chi$QSM)~\cite{Kim:2005gz} as well as the partial-wave analysis
for the non-strange pentaquark baryons~\cite{Azimov:2005jj}.
Moreover, a recent theoretical calculation of the $\gamma
N\rightarrow\eta N$ reaction~\cite{Choi:2005ki} describes
qualitatively well the GRAAL data, based on the values of the
transition magnetic moments in
Refs.~\cite{Kim:2005gz,Azimov:2005jj}, which implies that the
$N^{\ast}$ seen in the GRAAL experiment could be favorably
identified as one of the pentaquark baryons.

However, there is general opinion that, if the $\Theta^{+}$
exists, its width should be extremely small. Its value may even
possibly lie below 1 MeV~\cite{Arndt,Sibirtsev:2004bg}. As far as
theory is concerned, the decay width of the $\Theta^{+}$ has been
investigated in many different
approaches~\cite{Sibirtsev:2004cf,Cahn:2003wq,Karliner:2004qw,Hong:2004xn,
Ellis:2004uz,Praszalowicz:2004dn,Sibirtsev:2004bg,Walliser:2005pi}
and is mostly estimated to be very small. In the present work, we
want to study the decay width of the $\Theta^{+}$ baryon within
the framework of the chiral quark-soliton model ($\chi$QSM),
including the effects of flavor SU(3) symmetry breaking and using
the \textquotedblleft\emph{model-independent
approach}\textquotedblright~\cite{Adkins:1984cf}\footnote{The
approach is
\textquotedblleft\emph{model-independent}\textquotedblright~
 insofar that it does not perform self-consistent calculations
leading to some solitonic profile but uses only the semiclassical
rotational picture of the $\chi$QSM and determines the dynamical
coefficients by fitting them to experimental data. In fact the
pioneering paper on $\Theta^+$ \cite{Diakonov:1997mm} used this
method}. Recently, this approach has been applied to evaluate the
magnetic moments of the baryon decuplet and antidecuplet, with
parameters fixed by experimental magnetic moments of the baryon
octet~\cite{Yang:2004jr,Kim:1997ip} and the baryon octet, decuplet, 
and antidecuplet mass-splittings and the mass of the $\Theta^{+}$.
The same method was employed to get various transition magnetic 
moments~\cite{Kim:2005gz} and the results are in a good agreement
with the SELEX and GRAAL data. Thus, in the present work, we want
to analyze in the same way the axial-vector coupling constant of
the $\Theta^{+}$, based on the experimental data for hyperon
semileptonic decay (HSD) constants $(g_{1}
/f_{1})^{B_{1}\rightarrow B_{2}}$ and the flavor-singlet
axial-vector constant of the proton $g_{A}^{(0)}$. It is in
particular interesting to use the $g_{A}^{(0)}$ as an input, since
it carries information on the quark spin content of the proton. It
is extracted from deep inelastic polarized electron-proton
scattering and hence its information is independent of HSD.  In 
fact, the $g_{A}^{(0)}$ is related to the pseudoscalar coupling
$G_{2}$ in Ref.~\cite{Diakonov:1997mm} by the Goldberger-Treiman
relation but there it has been neglected, its effect being assumed
to be rather small.  In Ref.~\cite{Diakonov:1997mm} the
$\Gamma_{\Theta KN}$ was determined by the empirical value of the
pion-nucleon coupling constant $g_{\pi NN}$~\cite{Diakonov:2004ai}.
Moreover, the effects of flavor SU(3) symmetry breaking were
neglected. 

In the present work, we will perform a more general analysis of
the $\Gamma_{\Theta KN}$, emphasizing its dependence on
$g_{A}^{(0)}$ of the proton and the effects of SU(3) symmetry
breaking.  While the decay constants of HSDs are relatively well
determined~\cite{Yao:2006px}, the value of the $g_{A} ^{(0)}$ is
experimentally only known to be in the range of $0.15-0.35$
~\cite{Bass:2004xa}.  Thus, we need to examine explicitly the
dependence of the $\Gamma_{\Theta KN}$ on the $g_{A}^{(0)}$. We
will later show that the $\Gamma_{\Theta KN}$ is rather sensitive
to $g_{A}^{(0)}$ which is in contrast to the assumptions made in
Ref.~\cite{Diakonov:1997mm}. Moreover, we will see that the
$\Gamma_{\Theta KN}$ is constrained by the value of the
$\Sigma_{\pi N}$. In the end, we will see that the known data on 
HSDs and the experimental value of the $g_{A}^{(0)}$ are compatible
with a small width of $\Gamma_{\Theta KN} \leq 1$~MeV. 

\section{Formalism}
Using a formalism similar to the one of ref. \cite{Kim:2005gz} the
form factors of HSDs are defined by the following transition
matrix elements of the vector and axial-vector currents:
\begin{eqnarray}
\langle B_{2}|V_{\mu}^{X}|B_{1}\rangle&=&\bar{u}_{B_{2}}(p_{2})\left[
f_{1}(q^{2})\gamma_{\mu}-\frac{if_{2}(q^{2})}{M_{1}}\sigma_{\mu\nu}q^{\nu
}+\frac{f_{3}(q^{2})}{M_{1}}q_{\mu}\right]  u_{B_{1}}(p_{1}),\cr\langle
B_{2}|A_{\mu}^{X}|B_{1}\rangle&=&\bar{u}_{B_{2}}(p_{2})\left[  g_{1}
(q^{2})\gamma_{\mu}-\frac{ig_{2}(q^{2})}{M_{1}}\sigma_{\mu\nu}q^{\nu}
+\frac{g_{3}(q^{2})}{M_{1}}q_{\mu}\right]  \gamma_{5}u_{B_{1}}(p_{1}),
\end{eqnarray}
where the vector and axial-vector currents are defined as
\begin{equation}
V_{\mu}^{X}\;=\;\bar{\psi}(x)\gamma_{\mu}\lambda_{X}\psi(x),\;\;\;A_{\mu}
^{X}\;=\;\bar{\psi}(x)\gamma_{\mu}\gamma_{5}\lambda_{X}\psi
(x)\label{Eq:current}
\end{equation}
with $X=\frac{1}{2}(1\pm i2)$ for strangeness conserving $\Delta S=0$ currents
and $X=\frac{1}{2}(4\pm i5)$ for $|\Delta S|=1$. The $q^{2}=-Q^{2}$ stands for
the square of the momentum transfer $q=p_{2}-p_{1}$. The form factors $g_{i}$
and $f_{i}$ are real quantities due to $CP$-invariance, depending only on the
square of the momentum transfer. We can safely neglect $g_{3}$ for the reason
that its contribution to the decay rate is proportional to the ratio
$\frac{m_{l}^{2}}{M_{1}^{2}}\ll1$, where $m_{l}$ represents the mass of the
lepton ($e$ or $\mu$) in the final state and $M_{1}$ that of the baryon in the
initial state. Taking into account the $1/N_{c}$ rotational and $m_{\mathrm{s}
}$ corrections, we can write the resulting axial-vector constants
$g_{1}^{(B_{1}\rightarrow B_{2})}(0)$ as follows:
\begin{align}
g_{1}^{(B_{1}\rightarrow B_{2})}(0) &  =a_{1}\langle B_{2}|D_{X3}^{(8)}
|B_{1}\rangle\;+\;a_{2}d_{pq3}\langle B_{2}|D_{Xp}^{(8)}\,\hat{J}_{q}
|B_{1}\rangle\;+\;\frac{a_{3}}{\sqrt{3}}\langle B_{2}|D_{X8}^{(8)}\,\hat
{J}_{3}|B_{1}\rangle\nonumber\\
&  +m_{s}\left[  \frac{a_{4}}{\sqrt{3}}d_{pq3}\langle B_{2}|D_{Xp}
^{(8)}\,D_{8q}^{(8)}|B_{1}\rangle+a_{5}\langle B_{2}|\left(  D_{X3}
^{(8)}\,D_{88}^{(8)}+D_{X8}^{(8)}\,D_{83}^{(8)}\right)  |B_{1}\rangle\right.
\nonumber\\
&  +\left.  a_{6}\langle B_{2}|\left(  D_{X3}^{(8)}\,D_{88}^{(8)}-D_{X8}
^{(8)}\,D_{83}^{(8)}\right)  |B_{1}\rangle\right]  ,\label{Eq:g1}
\end{align}
where $a_{i}$ denote parameters encoding the specific dynamics of the chiral
soliton model. $\hat{J}_{q}$ ($\hat{J}_{3}$) stand for the $q$-th (third)
components of the collective spin operator of the baryons, respectively. The
$D_{ab}^{(\mathcal{R})}$ denote the SU(3) Wigner matrices in representation
$\mathcal{R}$.

The collective Hamiltonian describing baryons in the SU(3) $\chi$QSM takes the
following form \cite{Blotz:1992pw}:
\begin{equation}
\hat{H}=\mathcal{M}_{sol}+\frac{J(J+1)}{2I_{1}}+\frac{C_{2}(\text{SU(3)}
)-J(J+1)-\frac{N_{c}^{2}}{12}}{2I_{2}}+\hat{H}^{\prime}
\end{equation}
with the symmetry breaking piece given by:
\begin{equation}
\hat{H}^{\prime}=\alpha D_{88}^{(8)}+\beta Y+\frac{\gamma}{\sqrt{3}}
D_{8i}^{(8)}\hat{J}_{i}, \label{Hsplit}
\end{equation}
where parameters $\alpha$, $\beta$, and $\gamma$ are proportional to the
strange current quark mass $m_{s}$.

Taking into account the recent experimental observation of the mass of
$\Theta^{+}$, the parameters in Eq.(\ref{Hsplit}) can be conveniently
parameterized in terms of the pion-nucleon $\Sigma_{\pi N}$ term (assuming
$m_{s}/(m_{u}+m_{d})=12.9$) as \cite{Praszalowicz:2004dn}:
\begin{equation}
\alpha=336.4-12.9\,\Sigma_{\pi N},\quad\beta=-336.4+4.3\,\Sigma_{\pi N}
,\quad\gamma=-475.94+8.6\,\Sigma_{\pi N}\label{albega}
\end{equation}
(in units of MeV). Moreover, the inertia parameters which describe the
splittings of SU(3) baryon mass representations take the following values (in
MeV)
\begin{equation}
\frac{1}{I_{1}}=152.4,\quad\frac{1}{I_{2}}=608.7-2.9\,\Sigma_{\pi
N}.\label{ISigma}
\end{equation}
Equations (\ref{albega}) and (\ref{ISigma}) follow from the fit to the
masses of octet and decuplet baryons and of $\Theta^{+}$ as well. If
one imposes the additional constraint that
$M_{\Xi_{\overline{10}}}=1860$ MeV, then $\Sigma_{\pi N}=73$ MeV
\cite{Praszalowicz:2004dn} (see also
\cite{Schweitzer:2003fg}) in agreement with recent experimental
estimates~\cite{Pavan:2001wz,Inoue:2003bk}. However, since
the measurement of $M_{\Xi_{\overline{10}}}$ is still under debate, we
will not fix $\Sigma_{\pi N}$ but vary it within a certain range,
i.e. $\Sigma_{\pi N}=45-75$ MeV.

Because the Hamiltonian of Eq.(\ref{Hsplit}) mixes different SU(3)
representations, the collective wave functions are given as linear
combinations \cite{Kim:1998gt}:
\begin{eqnarray}
\label{admix}
\left\vert B_{8}\right\rangle  & =& \left\vert 8_{1/2}
,B\right\rangle +c_{\overline{10}}^{B}\left\vert \overline{10}_{1/2}
,B\right\rangle +c_{27}^{B}\left\vert 27_{1/2},B\right\rangle ,\cr
\left\vert B_{\overline{10}}\right\rangle &=&\left\vert \overline{10}
_{1/2},B\right\rangle +d_{8}^{B}\left\vert 8_{1/2},B\right\rangle
+d_{27} ^{B}\left\vert 27_{1/2},B\right\rangle
+d_{\overline{35}}^{B}\left\vert \overline{35}_{1/2},B\right\rangle,
\end{eqnarray}
where $\left\vert B_{\mathcal{R}}\right\rangle $ denotes the state which
reduces to the SU(3) representation $\mathcal{R}$ in the formal limit
$m_{s}\rightarrow0$. The spin index $J_{3}$ has been suppressed. The $m_{s}
$-dependent (through the linear $m_{s}$ dependence of $\alpha$, $\beta$, and
$\gamma$) coefficients in Eq.(\ref{admix}) can be found in
Ref.\cite{Yang:2004jr}.

\begin{table}[ht]
\begin{tabular}
[c]{l|lr}\hline
Decay modes & Experiments & Refs.\\\hline
${g_{1}}/{f_{1}}(n\to p)$ & $1.2695\pm0.0029$ & \cite{Yao:2006px}\\
${g_{1}}/{f_{1}}{(\Lambda\to p)}$ & $0.718\pm0.015$ & \cite{Yao:2006px}\\
${g_{1}}/{f_{1}}{(\Sigma^{-}\to n)}$ & $-0.34\pm0.017$ & \cite{Yao:2006px}\\
${g_{1}}/{f_{1}}{(\Xi^{-}\to\Lambda)}$ & $0.25\pm0.05$ & \cite{Yao:2006px}\\
${g_{1}}/{f_{1}}{(\Xi^{0}\to\Sigma^{+})}$ & $1.32_{-0.17}^{+0.21}\pm0.05$ &
\cite{Alavi-Harati:2001xk,Yao:2006px}\\
$g_{A}^{(0)}$ & $0.2 - 0.4$ & \cite{Goto:1999by}\\\hline
\end{tabular}
\caption{Experimental inputs for determining the dynamical parameters
  $a_{i}$.} 
\label{tab:input}
\end{table}

In order to determine the dynamical parameters $a_{i}$ in
Eq.(\ref{Eq:g1}), we want to use experimental information on HSD,
as we did for the magnetic moments of the SU(3)
baryons~\cite{Yang:2004jr}. Since there exist five experimental
data for the transition axial-vector coupling constants
$g_{1}/f_{1}(B_{1}\rightarrow B_{2})$ as listed in
Table~\ref{tab:input}, we need at least one more data. Since the
singlet axial-vector constant of the proton $g_{A}^{(0)}$ provides
an independen information from deep inelastic scattering of the
proton, we can use it for input as well. However, it has still a
large uncertainty, so that we will examine judicially the
dependence of the present analysis on $g_{A}^{(0)}$. With these
six input paramters at hand, the dynamical paramters $a_i$ can be
determined by solving the following matrix equation:
\begin{eqnarray}
\mathbf{M}[\Sigma_{\pi N}]\left[\begin{array}{c}
a_{1}\\
a_{2}\\
a_{3}\\
a_{4}\\
a_{5}\\
a_{6}\end{array}\right] & = & \left[\begin{array}{c}
1.2695\\
0.718\\
-0.34\\
0.25\\
1.32\\
0.2- 0.4
\end{array}\right]
\label{eq:matrixeq}
\end{eqnarray}
where
{\footnotesize
 \begin{eqnarray}
\mathbf{M}[\Sigma_{\pi N}] & = &
\left[\begin{array}{cccccc}
{\displaystyle
  -\frac{7}{15}-\frac{4c_{27}}{45}-\frac{2c_{\overline{10}}}{3}} &
{\displaystyle
  \frac{7}{30}-\frac{8c_{27}}{45}-\frac{2c_{\overline{10}}}{3}} &
{\displaystyle
  \frac{1}{30}+\frac{2c_{27}}{15}-\frac{c_{\overline{10}}}{3}} &
{\displaystyle -\frac{11}{135}} & {\displaystyle -\frac{2}{9}} &
{\displaystyle -\frac{2}{15}}\\
\\{\displaystyle
  -\frac{4}{15}+\frac{c_{27}}{15}+\frac{c_{\overline{10}}}{3}} &
{\displaystyle
  \frac{2}{15}+\frac{2c_{27}}{15}+\frac{c_{\overline{10}}}{3}} &
{\displaystyle
  \frac{1}{15}-\frac{c_{27}}{10}+\frac{c_{\overline{10}}}{6}} &
{\displaystyle -\frac{2}{45}} & {\displaystyle 0} & {\displaystyle
  -\frac{1}{15}}\\ \\
{\displaystyle \frac{2}{15}-\frac{2c_{27}}{45}} & {\displaystyle
  -\frac{1}{15}-\frac{4c_{27}}{45}} & {\displaystyle
  \frac{2}{15}+\frac{c_{27}}{15}} & {\displaystyle \frac{1}{135}} &
{\displaystyle -\frac{2}{45}} & {\displaystyle \frac{1}{15}}\\ \\
{\displaystyle -\frac{1}{15}-\frac{c_{27}}{15}} & {\displaystyle
  \frac{1}{30}-\frac{2c_{27}}{15}} & {\displaystyle
  \frac{1}{10}+\frac{c_{27}}{10}} & {\displaystyle \frac{1}{90}} &
{\displaystyle \frac{1}{15}} & {\displaystyle -\frac{1}{15}}\\ \\
{\displaystyle
  -\frac{7}{15}+\frac{2c_{27}}{45}+\frac{c_{\overline{10}}}{3}} &
{\displaystyle
  \frac{7}{30}+\frac{4c_{27}}{45}+\frac{c_{\overline{10}}}{3}} &
{\displaystyle
  \frac{1}{30}-\frac{c_{27}}{15}+\frac{c_{\overline{10}}}{6}} &
{\displaystyle -\frac{11}{270}} & {\displaystyle \frac{1}{9}} &
{\displaystyle \frac{1}{15}}\\ \\
{\displaystyle 0} & {\displaystyle 0} & {\displaystyle 1} &
{\displaystyle 0} & {\displaystyle -\frac{1}{5}} & {\displaystyle
  \frac{1}{5}}\end{array}\right].
\end{eqnarray} }
Inverting Eq.(\ref{eq:matrixeq}), we finally obtain the values of
dynamical parameters $a_i$ as functions of $\Sigma_{\pi N}$ and
$g_{A}^{(0)}$.

\section{Results and Discussion}
Table~\ref{tab:3} lists as example the results of the dynamical
parameters $a_i$ for $g_{A}^{(0)}=0.3$. The $a_i$ depend in
general on $g_{A}^{(0)}$ and $\Sigma_{\pi N}$ nonlinearly, except
for $a_6$ which is independent of $g_{A}^{(0)}$ as well as of
$\Sigma_{\pi N}$. As a matter of fact, the $a_i$ are generally
weakly dependent on $\Sigma_{\pi N}$ and the $a_2$, $a_4$, $a_5$
are rather sensitive to $g_{A}^{(0)}$.

\begin{table}[ht]
  \begin{tabular}{c|cccccc}
\hline
$\Sigma_{\pi N}[{\rm MeV}]$&
{\large $a_{1}$}&
{\large $a_{2}$}&
{\large $a_{3}$}&
{\large $a_{4}$}&
{\large $a_{5}$}&
{\large $a_{6}$}\tabularnewline
\hline
$45$&
$-2.4811$&
$0.8933$&
$0.3190$&
$1.3580$&
$0.1399$&
$0.0450$\tabularnewline
$50$&
$-2.4113$&
$1.0303$&
$0.3196$&
$1.3464$&
$0.1432$&
$0.0450$\tabularnewline
$55$&
$-2.3608$&
$1.1255$&
$0.3210$&
$1.3226$&
$0.1499$&
$0.0450$\tabularnewline
$60$&
$-2.3221$&
$1.1952$&
$0.3228$&
$1.2904$&
$0.1591$&
$0.0450$\tabularnewline
$65$&
$-2.2909$&
$1.2481$&
$0.3250$&
$1.2517$&
$0.1700$&
$0.0450$\tabularnewline
$70$&
$-2.2650$&
$1.2895$&
$0.3275$&
$1.2076$&
$0.1825$&
$0.0450$\tabularnewline
$75$&
$-2.2426$&
$1.3224$&
$0.3303$&
$1.1587$&
$0.1964$&
$0.0450$\tabularnewline
\hline
  \end{tabular}
\caption{The dynamical parameters $a_i$ determined with
  $g_{A}^{(0)}=0.3$. The $\Sigma_{\pi N}$ is varied
  from $45$ to $75$ MeV.}
\label{tab:3}
\end{table}

We now insert the values of $a_i$ listed in Table~\ref{tab:3} and
similar ones with different $g_{A}^{(0)}$  into Eq.(\ref{Eq:g1})
together with the matrix elements of the $D_{ab}^{(\mathcal{R})}$
functions so that we can determine the transition axial-vector
constant for the decay $\Theta^+\to K^+ n$.
\begin{figure}[ht]
\centering
\includegraphics[scale=0.6]{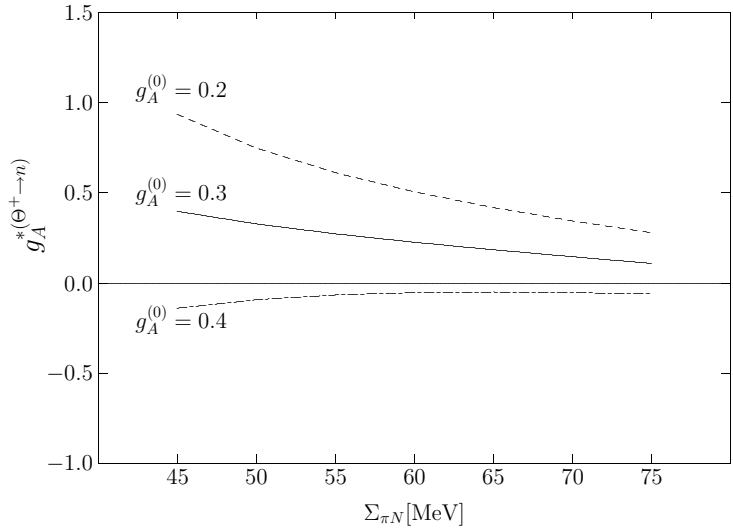}
\caption{The transition axial-vector coupling constant for
  $\Theta^+\to K^{+} n$ as a function of $\Sigma_{\pi N}$.  The solid
  curve denotes that with $g_A^{(0)}=0.3$,
  while the dashed and dot-dashed ones represent that with
  $g_A^{(0)}=0.2,\;0.4$, respectively.}
\label{fig:1}
\end{figure}
Fig.~\ref{fig:1} draws for various values of $g_{A}^{(0)}$ the
dependence of the $g_A^{*(\Theta\to n)}$ on the $\Sigma_{\pi N}$.
The larger $g_{A}^{(0)}$ we use, the smaller $g_A^{*(\Theta\to
n)}$ we obtain. Moreover, the $g_A^{*(\Theta\to n)}$ turns out to
be negative when $g_{A}^{(0)}$ is larger than around $0.37$. For a
given $g_{A}^{(0)}$ the $g_A^{*(\Theta\to n)}$ decreases as the
$\Sigma_{\pi N}$ increases. The $g_A^{*(\Theta\to n)}$ at
$\Sigma_{\pi N}=70$ MeV is 70 \% smaller than that at $\Sigma_{\pi
N}=45$ MeV.

Using the effective Lagrangian for the
$\Theta^+ \to K^{+} n$ decay:
\begin{equation}
\mathcal{L}  =  -\frac{g_{A}^{\ast(\Theta\to n)}}
{2f_{K}}\overline{\Theta} \gamma_{\mu}\gamma_{5}
\left(\partial^{\mu}K^{+}\right)n + \mathrm{h.c.},
\label{eq:elag}
\end{equation}
where $\mathrm{h.c.}$ stands for the hermitian conjugate.  We obtain
from the effective Lagrangian the invariant amplitude for the decay as
follows:  
\begin{equation}
i\mathcal{M}  =  i\frac{g_{A}^{\ast(\Theta\to n)}}
{2f_{K}}\overline{u}(p_{n})p_{\mu}^{K^+}\gamma_{5}
\gamma^{\mu}u(p_{\Theta}),
\label{eq:damp}
\end{equation}
where $\overline{\Theta}$, $K^+$, and $n$ denote the fields of the
$\Theta^+$, of the positively charged kaon, and of the neutron,
respectively. The $f_K=112~\rm{MeV}$ represents the kaon decay
constant. The $\overline{u}(p_n)$ and $u(p_\Theta)$ are the Dirac
spinors for the neutron and $\Theta^+$ with the corresponding
momenta, respectively and the $p^{K^+}$ denotes the kaon momentum.
The decay width of the $\Theta^+\to KN$ is proportional to the
square of the transition axial-vector constant:
\begin{equation}
\Gamma_{\Theta KN}  = 2\Gamma_{\Theta K^{+} n} =
 \frac{\left(g_{A}^{\ast(\Theta\to n)}\right)^{2} |\vec{p}|}{16\pi
 f_{K}^{2}M_{\Theta}^{2}}\left[\left(M_{\Theta}-M_{N}
\right)^{2}-m_{K}^{2}\right]\left(M_{\Theta}+M_{N}\right)^{2},
\end{equation}
where $|\vec{p}|=\sqrt{(M_{\Theta} ^{2}
-(M_{N}+m_{K})^2)(M_{\Theta}^{2} -(M_{N}-m_{K})^2)} / 2M_{\Theta}$
is the kaon momentum and $M_\Theta =1540~\rm{MeV}$, $M_N
=939~\rm{MeV}$, and $m_K =494~\rm{MeV}$ stand for the masses of
the $\Theta^+$, the nucleon, and the kaon, respectively. The
factor 2 in front of the decay width $2 \Gamma_{\Theta K^{+} n}$
takes care of the fact that $\Theta^+$ has two distinct decay
channels, $K^{+} n$ and $K^{0} p$, which are equally populated due
to isospin-symmetry. The $\Gamma_{\Theta KN}$ is sensitive to the
value of the $g_A^{*(\Theta \to n)}$. Moreover, since the
$\Gamma_{\Theta KN}$ is given as a function of the square of the
$g_A^{*(\Theta\to n)}$, it is independent of the sign of the
$g_A^{*(\Theta\to n)}$. Thus, the decay width $\Gamma_{\Theta KN}$
decreases until the sign of the $g_A^{*(\Theta\to n)}$ changes and
then increases again.

\begin{table}[ht]
\begin{tabular}{c|cccc}
\hline
\multicolumn{1}{c|}{$\Gamma_{\Theta KN}^{({\rm total})}$}&
\multicolumn{4}{c}{Input $g_{A}^{(0)}$}\\
\textbf{\small $\Sigma_{\pi N}[{\rm MeV}]$}& \textbf{\small
~~~~$0.28$~~~~}& \textbf{\small ~~~~$0.32$~~~~}& \textbf{\small
~~~~$0.36$~~~~}& \textbf{\small ~~~~$0.40$~~~~} \\ \hline $45$&
$33.41$& $11.06$& $0.76$&
$2.51$  \\
$50$& $22.25$& $7.82$& $0.76$&
$1.10$\\
$55$& $15.22$& $5.53$& $0.64$&
$0.56$\\
$60$& $10.45$& $3.82$& $0.46$&
$0.36$\\
$65$& $7.04$& $2.51$& $0.26$&
$0.31$\\
$70$& $4.54$& $1.50$& $0.10$&
$0.35$\\
$75$& $2.70$& $0.75$& $0.01$&
$0.47$\\
\hline
\end{tabular}
\caption{The decay width of $\Theta^+\to K N$ determined with
$g_{A}^{(0)}$ varied from $0.28$ to $0.40$. The $\Sigma_{\pi N}$
is varied from $45$ to $75$ MeV.} \label{tab:results}
\end{table}
In Table~\ref{tab:results}, we list the total decay width of the
$\Theta^+\to K N$ as a function of $\Sigma_{\pi N}$ and
$g_{A}^{(0)}$. Actually, the decay width decreases until
$g_{A}^{(0)}=0.4$, and then starts to increase, while it gets
smaller almost monotonically as the larger value of the
$\Sigma_{\pi N}$ is used. The region where proper combinations of
$g_{A}^{(0)}$ and $\Sigma_{\pi N}$ yield a small width
$\Gamma_{\Theta KN} \leq 1$~MeV of the $\Theta^+$ pentaquark can
easily be identified.

\begin{figure}[ht]
\centering
\includegraphics[scale=0.8]{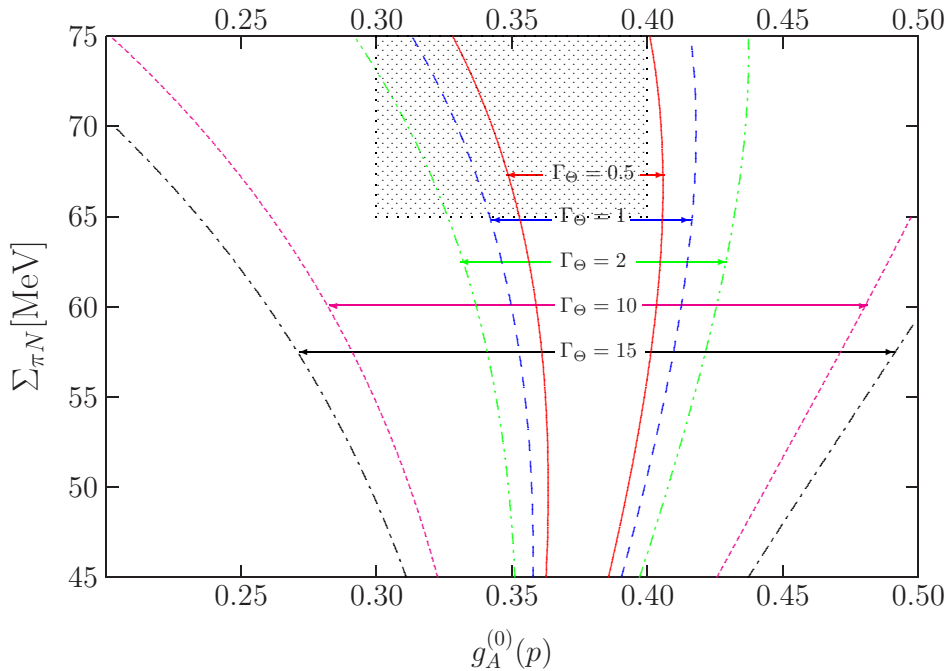}

\caption{The total decay width of $\Theta^+\to K N$ in units of {\rm MeV}
  as a function of $g_{A}^{(0)}$ and $\Sigma_{\pi N}$.  The shaded
  square denotes the ranges of $g_{A}^{(0)}$: $0.3- 0.4$ and of
  $\Sigma_{\pi N}$: $65 - 75$ MeV. }
\label{fig:2}
\end{figure}

In Fig.~\ref{fig:2} we draw the results of the total decay width
of the $\Theta^+\to K N$ as a function of $\Sigma_{\pi N}$ and
$g_{A}^{(0)}$. The smaller the $\Gamma_{\Theta KN}$ the more
restricted are the values $\Sigma_{\pi N}$ and $g_{A}^{(0)}$. The
shaded rectangle indicates the area where one has generally
accepted experimental values of $g_{A}^{(0)}$ and  $\Sigma_{\pi N}$, 
i.e. $0.3- 0.4$ and $65 - 75$ MeV, respectively, 
and simultaneously a $\Gamma_{\Theta KN} \leq 1$ MeV.  It is of great 
interest to see that the range of $g_{A}^{(0)}$ is compatible with a
theoretical investigation~\cite{Wakamatsu:2006ba}, based on the
$\chi$QSM, on the COMPASS and HERMES measurements of the deuteron
spin-dependent structure function~\cite{Ageev:2005gh,
Alexakhin:2006vx, Airapetian:2006vy}. It is worthwhile to mention that
the values of $g_A^{(0)}$ in the present analysis is almost the
same as theoretical results within the
$\chi$QSM~\cite{Wakamatsu:1999nj,Silva:2005fa}.  The range of
$\Sigma_{\pi N}$ given above is consistent with a recent
analysis~\cite{Schweitzer:2003fg}.  If one interprets the result of
the DIANA collaboration~\cite{Barmin:2006we} as identification of the
$\Theta^+$, namely the formation of a narrow $pK^{0\text{ }}$ peak
with mass of $1537\pm2$ MeV/$c^{2}$ and width of $\Gamma=0.36\pm0.11$
MeV in the $K^{+}n\rightarrow K^{0}p$ transition, then that result is
inside the shaded area of Fig.~\ref{fig:2}.  

\section{Summary and Conclusion}
In the present work, we analyzed within the framework of the
chiral quark-soliton model the total decay width of the
$\Theta^+\to KN$, based on the experimental data of hyperon
semileptonic decays and the flavor-singlet axial-vector constant 
$g_{A}^{(0)}$. The parameters in the collective Hamiltonian were
fixed by the splittings of the SU(3) baryon mass
representation~\cite{Praszalowicz:2004dn}. The dynamical
parameters in the collective axial-vector operators were fitted
(\textquotedblleft\emph{model-independent
approach}\textquotedblright) to the existing data of hyperon
semileptonic decays and of $g_{A}^{(0)}$, where the value of the
$g_{A}^{(0)}$ was varied within the range of $0.2- 0.5$. Since all
these parameters depend on the value of the $\Sigma_{\pi N}$, we
took its value to be $45- 75$ MeV.

We first computed the transition axial-vector coupling constant
for the $\Theta^+\to K^{+} n$, $g_A^{*(\Theta\to n)}$. We showed
that the $g_A^{*(\Theta\to n)}$ decreases as $g_{A}^{(0)}$
increases. Furthermore, the $g_A^{*(\Theta\to n)}$ depends on the
$\pi N$ sigma term, $\Sigma_{\pi N}$: it is getting smaller as the
$\Sigma_{\pi N}$ increases. Thus, the $g_A^{*(\Theta\to n)}$ turns
out to be smaller with $\Sigma_{\pi N}=70$ MeV by 70 \%, compared
to that with $\Sigma_{\pi N}=45$ MeV. It was also found that the
$g_A^{*(\Theta\to n)}$ becomes negative around $g_{A}^{(0)}\simeq
0.37$.

The total width $\Gamma_{\Theta KN}$ of the $\Theta^+\to KN$ decay
was finally investigated. Since it is proportional to the square
of the transition axial-vector constant $g_A^{*(\Theta\to n)}$, it
is rather sensitive to the $g_A^{*(\Theta\to n)}$. The
$\Gamma_{\Theta KN}$ is getting suppressed as the singlet
axial-vector constant $g_{A}^{(0)}$ increases. However, since the
$g_A^{*(\Theta\to n)}$ turns out to be negative around 0.37, the
$\Gamma_{\Theta KN}$ starts to increase around 0.37. As a result,
the total decay width $\Gamma_{\Theta KN}$ turns out to be smaller
than 1 MeV for values of the $g_{A}^{(0)}$ and $\Sigma_{\pi N}$
larger than 0.31 and 65 MeV, respectively.

As conclusion of present analysis, which uses the
\textquotedblleft\emph{model-independent approach} to the chiral quark
soliton, one can state: The known data of semileptonic decays
combined with $0.3 \leq  g_{A}^{(0)} \leq 0.4$ and $\Sigma_{\pi N}
\geq 65$ MeV is compatible with the existence of a $\Theta^+$
pentaquark having a small width of the total decay $\Theta^+\to KN$:
$\Gamma_{\Theta KN}\le 1~ \rm{MeV}$.  Since all dynamical parameters
in the present approach are fitted the existing experimental data, it
is difficult to understand the origin of the strong correlation
between the singlet axial-vector constant and the $\Theta^+$ decay
width.  The corresponding investigation is under way, in order to give
a theoretical explanation of this strong
correlation~\cite{Ledwigetal}.

\section*{Acknowledgments}
The authors are grateful to J.K. Ahn, D. Diakonov, J.H. Lee,
S.i. Nam, M.V. Polyakov, and M. Prasza\l owicz for helpful discussion
and comments.  The present work is supported by the Korea
Research Foundation Grant funded by the Korean Government(MOEHRD)
(KRF-2006-312-C00507).  The work is also supported by the
Transregio-Sonderforschungsbereich Bonn-Bochum-Giessen, the
Verbundforschung (Hadrons and Nuclei) of the Federal Ministry for
Education and Research (BMBF) of Germany, the Graduiertenkolleg
Bochum-Dortmund, the COSY-project J\"ulich as well as the EU
Integrated Infrastructure Initiative Hadron Physics Project
under contract number RII3-CT-2004-506078.

\end{document}